\documentclass[aps,prd,twocolumn,showpacs,showkeys,amsmath,amssymb]{revtex4}
\usepackage{graphicx}
\usepackage{bm}
\begin{document}

\title{The Narrow $\Theta_5$ Pentaquark As
The First Non-planar Hadron With the Diamond Structure And
Negative Parity}
\author{Xing-Chang Song}
\affiliation{%
Department of Physics, Peking University, BEIJING 100871, CHINA}
\author{Shi-Lin Zhu}
\email{zhusl@th.phy.pku.edu.cn}
\affiliation{%
Department of Physics, Peking University, BEIJING 100871, CHINA}
\affiliation{%
COSPA, Department of Physics, National Taiwan University, Taipei
106, Taiwan, R.O.C.}

\date{\today}

\begin{abstract}
Using the picture of the flux tube model, we propose that the
$\Theta_5$ pentaquark as the first candidate of the
three-dimensional non-planar hadron with the extremely stable
diamond structure. The up and down quarks lie at the corners of
the diamond while the anti-strange quark sits in the center.
Various un-excited color flux tubes between the five quarks bind
them into a stable and narrow color-singlet. Such a configuration
allows the lowest state having the negative parity naturally. The
decay of the $\Theta_5$ pentaquark into the nucleon and kaon
requires the breakup of the non-planar diamond configuration into
two conventional planar hadrons, which involves some kind of
structural phase transition as in the condensed matter physics.
Hence the width of the $\Theta^+$ pentaquark should be narrow
despite that it lies above the kaon nucleon threshold. We suggest
that future lattice QCD calculation adopt non-planar interpolating
currents to explore the underlying structure of the $\Theta_5$
pentaquark.
\end{abstract}

\pacs{12.39.Mk, 12.39.-x}

\keywords{Pentaquark, Diamond Structure}

\maketitle

\pagenumbering{arabic}

\section{Introduction}\label{sec1}

Up to now nine experimental Collaborations reported evidence for
the existence of the narrow $\Theta^+$ pentaquark state with the
minimal quark content $uudd\bar s$ around $1540$ MeV \cite{leps}.
The third component of its isospin is $I_z=0$ while its total
iso-spin is likely $I=0$. Its angular momentum and parity have not
been determined from experiments. Later, NA49 Collaboration
\cite{na49} reported another narrow pentaquark candidate
$\Xi_5^{--}$ with strangeness $S=-2$, baryon number $B=1$, and
isospin $I=\frac32$ around $1862$ MeV. However the existence of
this state is still under debate \cite{doubt} and confirmation
from other groups is necessary.

These exotic baryons are clearly beyond the conventional quark
model, which has been successful in the classification of hadron
states although its foundation has not been derived from Quantum
Chromo-dynamics (QCD). Non-conventional quark model states such as
glueballs, hybrid mesons, other multi-qaurk hadrons are allowed in
the QCD spectrum, which have motivated the extensive experimental
searches for the past several decades. Unfortunately none of them
was pinned down without controversy \cite{pdg}. The emergence of
the $\Theta^+$ pentaquark finally unveiled one corner of the
curtain over the vast landscape of multi-quark hadrons, creating
both confusion and anticipation.

Theoretical study of pentaquark states dated back to the early
days of QCD using MIT bag model \cite{strot}. The recent interest
in the $\Theta^+$ pentaquark started with the prediction of its
mass, width and reaction channel from the chiral soliton model
(CSM) in Ref. \cite{diak} a few years ago. However this prediction
is very sensitive to the model inputs \cite{diak,mp,ellis}. For
example, either adopting the commonly used value $45$ MeV for the
$\sigma$-term or identifying $N(1710)$ as a member of the
anti-decuplet will lead to a $\Xi_5^{--}$ pentaquark mass $210$
MeV higher than that observed by NA49 Collaboration \cite{diak}.
With the new larger value $(79\pm 7)$ MeV and $(64\pm 7)$ MeV from
two recent analysis \cite{sigmaterm} for the $\sigma$-term, a
fairly good description of both $\Theta^+$ and $\Xi^{--}$ masses
is possible \cite{ellis}. In CSM, the narrow width comes from the
unnatural and accidental cancellation between the coupling
constants in the leading order, next-leading order and
next-next-leading order large $N_c$ expansion. Moreover, the
theoretical foundation of the treatment of the pentaquarks in the
chiral soliton model is challenged by the large $N_c$ formalism in
Refs. \cite{cohen}.

Since early last year, there appeared many theoretical papers
trying to interpret these exotic states
\cite{page,jaffe,zhu,shuryak,lipkin,buddy,huang,qsr,lattice,zhang,width1,width2,width3,yu}.
Several clustered quark models \cite{jaffe,shuryak,lipkin} were
constructed to ensure the pentaquarks have positive parity as in
the original chiral soliton model. But the $\Theta^+$ pentaquark
parity is still a pending issue. For example, one recent lattice
calculation favors positive parity for pentaquarks \cite{chiu}
while two previous lattice simulations favor negative parity in
Refs. \cite{lattice}. QCD sum rule approach also favors negative
parity \cite{zhu,qsr}. Some other models favor negative parity as
well \cite{zhang}. Recently, many theoretical papers proposed
interesting ways to determine $\Theta^+$ parity \cite{yu}.

Among these models, Jaffe and Wilczek's (JW) clustered diquark
model is a typical one \cite{jaffe}. They proposed that there
exists strong correlation between the light quark pair when they
are in the anti-symmetric color $({\bar 3}_c)$, flavor $({\bar
3}_f)$, isospin $(I=0)$ and spin $(J=0)$ configuration
\cite{jaffe}. The lighter the quarks, the stronger the
correlation, which helps the light quark pair form a diquark. For
example, the ud diquark behaves like a scalar with positive
parity. Such correlation may arise from the color spin force from
the one gluon exchange or the flavor spin force induced by the
instanton interaction. In order to accommodate the $\Theta^+$
pentaquark, Jaffe and Wilczek required the flavor wave function of
the diquark pair to be symmetric ${\bf {\bar 6_f}}$ and their
color wave function to be antisymmetric ${\bf 3}_c$. Bose
statistics of the scalar diquarks demands an odd orbital
excitation between the two diquarks, which ensures that the
resulting pentaquark parity is positive. The flavor anti-decuplet
is always accompanied by an octet which is nearly degenerate and
mixes with the decuplet.

Jaffe and Wilczek pointed out that one of the decay modes
$\Xi_5\to \Xi^\ast + \pi$ observed by NA49 \cite{na49} signals the
existence of an additional octet around $1862$ MeV together the
anti-decuplet since the latter can not decay into a decuplet and
an octet in the $SU(3)_f$ symmetry limit. If further confirmed,
this experiment poses a serious challenge to the chiral soliton
model because there is no baryon pentaquark octet in the
rotational band in this model. But it may be hard to exclude it
since there always exist excited vibrational octet modes. These
modes can not be calculated rigourously within the chiral soliton
model \cite{ellis}.

However, in JW's model $\Theta^+$ is not the lightest pentaquark
as in the chiral soliton model. The ideal mixing between the octet
and anti-decuplet will split the spectrum and produce two
nucleon-like states. The higher one $N_s$ is around 1710 MeV with
a quark content $qqq \bar s s$ where $q$ is the up or down quark.
The lighter one $N_l$ is lower than the $\Theta^+$ pentaquark with
a quark content $qqq \bar q q$. Jaffe and Wilczek identified $N_l$
as the well-known Roper resonance $N(1440)$ \cite{jaffe}, which is
a very broad four-star resonance with a width around $(250 \sim
450)$ MeV. As two members within the same anti-decuplet, it will
be very demanding to explain Roper's large decay width and
$\Theta^+$'s extremely narrow width simultaneously in a natural
way. Further analysis of the phenomenological constraints on this
model can be found in \cite{tomcohen}.

\section{The Narrow Width Puzzle}\label{sec2}

According to these experiments, both $\Theta^+$ and $\Xi^{--}_5$
are very narrow states. In fact, the $\Theta^+$ pentaquark is so
narrow that most of the experiments can only set an upper bound
around $20$ MeV. Recent analysis of kaon nucleon scattering data
indicates that the width of the $\Theta^+$ pentaquark is less than
several MeV \cite{nuss}, otherwise they should have shown up in
these old experiments.

Experience with conventional excited hadrons shows that their
widths are around one hundred MeV or even bigger if they lie 100
MeV above threshold and decay through S-wave or P-wave. For
comparison, the $S=-1$ hyperon $\Lambda (1520)$ $D_{03}$ state is
in the same mass region as the $\Theta^+$ pentaquark. Its angular
momentum and parity is $J^P={3\over 2}^-$. Its dominant two-body
decay is of D-wave with final states $N\bar K$, $\Sigma \pi$. With
a smaller phase space and higher partial wave, the width of
$\Lambda (1520)$ is $15.6\pm 1.0$ MeV \cite{pdg}. In contrast,
$\Theta^+$ lies above 100 MeV $K N$ threshold and decays through
either S-wave or P-wave with a total width less than several MeV,
corresponding to negative or positive parity respectively. If
these states are further established and confirmed to have such a
narrow width, the most challenging issue is to understand their
extremely narrow width in a natural way. Is there a mysterious
selection rule which is absent from the conventional hadron
interaction? This is the topic of the present paper.

Recently there have been several attempts to explain the narrow
width of the $\Theta^+$ pentaquark \cite{width1,width2,width3}.
Carlson et al. constructed a special pentaquark wave function
which is totally symmetric in the flavor-spin part and
anti-symmetric in the color-orbital part in Ref. \cite{width1}.
With this wave function they found that the overlap probability
between the pentaquark and the nucleon kaon system is ${5\over
96}$. Taking into account of the orbital wave function in JW's
diquark model further reduces the overlap probability to ${5\over
596}$ \cite{width1}. The small overlap probability might be
responsible for the narrow width of pentaquarks.

In Ref. \cite{width2} Karliner and Lipkin proposed that there
might exist two nearly generate pentaquarks. Both of them decay
into the kaon and nucleon. Hence these two states mix with each
other by the loop diagram via the decay modes. Diagonalization of
the mass matrix leads to a narrow $\Theta^+$ pentaquark which
almost decouples with the decay mode. The pentaquark with the same
quantum number is very broad with a width of around 100 MeV, which
has escaped the experimental detection so far.

In Ref. \cite{width3} Buccella and Sorba suggested that the four
quarks are in the $L=1$ state and the anti-quark is in the S-wave
state inside the $\Theta^+$ and $\Xi^{--}_5$ pentaquarks. When the
anti-quark picks up a quark to form a meson, two of the other
three quarks remain in the $SU(6)_{fS}$ totally anti-symmetric
state which is orthogonal to that of the $SU(6)_{FS}$ totally
symmetric representation for the nucleon octet. This selection
rule is exact in the $SU(3)_f$ symmetry limit. The narrow widths
of the $\Theta^+$ and $\Xi^{--}_5$ pentaquarks come from the
$SU(3)_f$ symmetry breaking.

\section{The Diamond Structure For The Narrow $\Theta_5$ Pentaquarks}\label{sec4}

In this note we propose an alternate scheme to explain the narrow
width of the $\Theta^+$ pentaquark.

For the light quarks, their motion inside the hadrons is fully
relativistic. Talking about the spatial configuration may seem
misleading at the first sight. We deal with this issue in the
framework of the flux tube model \cite{flux}.

According to this model, the strong color field between a pair of
quark and anti-quark forms a flux tube which confines them. The
flux tube is not excited for the ordinary mesons. When it is
excited, this system appears as a hybrid meson. Similarly there
exist flux tubes between the quark pairs inside the baryon. Recent
lattice simulations of static baryon interaction tend to support
the flux tube picture \cite{flux-lattice}, although whether the
flux tube is of Y-shape or $\Delta$-Shape is still under debate.

It's interesting to note that all the conventional mesons and
baryons are planar hadrons. At every moment, if we could take a
picture of these hadrons, we would find that all the quarks and
flux tubes inside the hadrons lie on the same plane.

We propose the narrow $\Theta_5$ pentaquark as the first candidate
of the non-planar hadron with the diamond structure. The strange
anti-quark sits in the center while the up and down quarks lie at
the corners of the diamond. Four flux tubes between the $\bar s$
and four light (u or d) quarks bind the system into a color
singlet.

It is understood that the term "center" or "corner" denotes the
relative position which the corresponding quark will occupy with
the maximum probability. At every moment, if we could take a
real-time picture, we would see a centered diamond structure with
the maximum probability.

If the isospin symmetry is exact, the four quarks are exactly on
the same footing. For the ground state of this system, none of the
four flux tubes is excited. And none of the five quarks is
orbitally excited. The diamond structure is exact. So the parity
of the lowest state is negative. The isospin is zero.

It's well known that the diamond structure is extremely stable in
nature. When the $\Theta_5$ pentaquark decays into the planar kaon
and nucleon, this system undergoes a special structural phase
transition, breaking the non-planar flux tubes and forming new
planar ones. Hence the decay width of the $\Theta_5$ pentaquark
should be small.

Replacing the $\bar s$ by $\bar c$ and $\bar b$ we will arrive at
narrow heavy pentaquarks with the same diamond structure.
Replacing the $\bar s$ by $\bar u$ and $\bar d$ will cause
annihilation of the light quark pair, thus change the whole
picture. We do not discuss this case here.

For the $\Xi_5$ with quark content $uuss\bar d$, the diamond
structure is severely distorted because of explicit SU(3) symmetry
breaking. Now $\bar d$ sits in the center. Although $\Xi_5$ is
also a narrow state, its width is bigger than that of $\Theta_5$
since it is relatively easier to break the distorted diamond
structure.

We strongly urge that future lattice QCD simulations employ
non-local and non-planar interpolating currents to explore the
pentaquark structure. If the $\theta_5$ pentaquark really possess
the diamond structure, it may not couple to the local
interpolating current with five quark fields at the same point
very strongly because of the topology difference.

We note that Liu et al performed the inherent nodal structure
analysis to both the square and the equilateral terahedron based
on the pure symmetry consideration \cite{liuyx}. They concluded
that the parity of the $\Theta_5$ pentaquark is positive due to
the orbital excitation \cite{liuyx}, opposite to what we obtained
in the present work.

In short summary, we have proposed the diamond structure to
explain the narrow width of the pentaquarks. In the future, the
full three-dimensional $\Theta_5$ pentaquark structure may be
explored using the Wigner-type phase space distribution as
suggested in Ref. \cite{ji}.

If future experiments confirm the diamond structure for the
$\Theta_5$ pentaquark, we propose the next interesting candidate
of the non-planar hadron is the fullerene-like $\mbox{Quark}_{60}$
with sixty valence quarks and baryon number B=20. Such a state is
obtained by replacing the carbon atom in the $C_{60}$ by a valence
quark and the corresponding QED valence bonds by the QCD flux
tubes. Such a cage-like structure may ensure that this B=20
particle is very stable.

This project was supported by the National Natural Science
Foundation of China under Grant 10375003, Ministry of Education of
China, FANEDD and SRF for ROCS, SEM. S.L.Z. thanks Prof W.-Y. P.
Hwang and COSPA center at National Taiwan University for the warm
hospitality.


\end{document}